\documentclass[prl,twocolumn]{revtex4}%
\usepackage{amsmath}
\usepackage{amsfonts}
\usepackage{amssymb}
\usepackage{graphicx}%
\begin{document}


\title{Taming the Blackbody with Metamaterials}
	
\author{Xianliang Liu$^1$, Talmage Tyler$^2$, Tatiana Starr$^3$, Anthony F. Starr$^3$, Nan Marie Jokerst$^2$, and Willie J. Padilla$^1$}
\email[]{willie.padilla@bc.edu}

\affiliation{$^1$Department of Physics, Boston College, 140 Commonwealth Ave., Chestnut Hill, MA 02467, USA}
\affiliation{$^2$Department of Electrical and Computer Engineering, Duke University, Durham, North Carolina 27708, USA}
\affiliation{$^3$SensorMetrix, Inc., San Diego, California 92121, USA}

\begin{abstract}

In this paper we demonstrate, for the first time, selective thermal emitters based on metamaterials perfect absorbers. We experimentally realize a narrow band mid-infrared (MIR) thermal emitter. Multiple metamaterial sublattices further permit construction of a dual-band MIR emitter. By performing both emissivity and absorptivity measurements, we find that emissivity and absorptivity agree very well as predicted by Kirchhoff's law of thermal radiation. Our results directly demonstrate the great flexibility of metamaterials for tailoring blackbody emission.

\end{abstract}

\maketitle

A blackbody is an idealized object that absorbs all radiation incident upon it and re-radiates energy solely determined by its temperature, as described by Planck's law~\cite{Planck01}. In reality no object behaves like an ideal blackbody, instead the emitted radiation is determined by the absorption of the material. The desire to control radiated energy has long been a research topic of interest for scientists - one particular theme being the construction of a selective emitter whose thermal radiation is much narrower than that of a blackbody at the same temperature.

Various thermal emitters are desired for energy harvesting applications, such as in the field of thermophotovoltaics (TPVs), in that the conversion efficiently can be greatly enhanced. Extensive efforts have been made using the luminescent bands of rare earth oxides for selective emission~\cite{Chubb99,Bitnar02,Torsello04}. However such selective emitters are limited by the availability of materials and cannot be controlled beyond mixing various compounds, thus limiting their performance. Photonic crystals have been investigated as an alternative candidate for selective emitters~\cite{Pralle02,Lin03,Laroche00}. However due to their non-resonant nature photonic crystal emitters do not have very sharp bands or high emissivity and therefore do not significantly increase efficiencies.

\begin{figure}[!]
\begin{center}
\includegraphics[ width=2.75in,keepaspectratio=true]%
{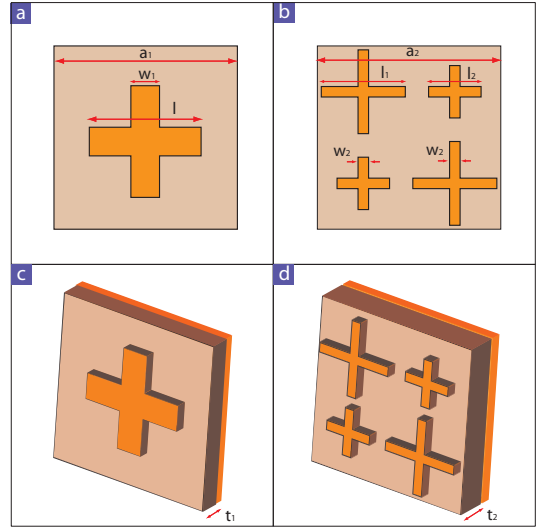}%
\caption{(Color online) Design of the infrared metamaterial absorber. a, Top view of a single band metamaterial absorber unit cell with dimensions of: a = 3.2, l = 1.7, w$_{1}$ = 0.5, in microns. b, Schematic of a dual band metamaterial absorber with dimensions, (in microns): a$_{2}$ = 7, l$_{1}$ = 3.2, l$_{2}$ = 2.0, w$_{2}$=0.4.  c, d, Perspective view for single and dual band metamaterial absorbers. Thickness of dielectric spacer is: t$_{1}$ = 0.2 $\mu$m, t$_{2}$ = 0.3 $\mu$m.}
\label{Fig1}%
\end{center}
\end{figure}

Another proposed technology which may be relevant for thermophotovoltaic applications is metamaterials~\cite{Fu09,Avitzour09,Padilla10} - artificial electromagnetic materials normally composed of periodic metallic elements. Metamaterials have demonstrated the ability to achieve exotic properties difficult to attain with nature materials. One extraordinary property explored early on in metamaterials research is negative refractive index - theoretically predicted in 1968~\cite{Veselago68} and experimentally demonstrated after 2000~\cite{Smith00,Schultz01,Smith04}. Since then, research into metamaterials has grown enormously resulting in many novel phenomena including invisibility cloaks~\cite{Schurig06} and perfect lenses~\cite{Pendry00, Fang05}. An intriguing use of metamaterials has been development of the so called `perfect absorber', which exhibits the ability to yield near-unity absorptivity in nearly any frequency range~\cite{Landy08,Tao082,Tao08,Liu10,Hao10}. According to Kirchhoff's law of thermal radiation, at equilibrium the emissivity of a material equals its absorptivity. Therefore in principle, metamaterial perfect absorbers radiate energy as described by their absorptivity, at a given temperature. Due to the resonant nature of metamaterials the perfect absorber yields sharp resonances with high absorption, thus suggesting their use as high-Q emitters with high emissivity. Here we report on selective thermal emitters in the mid-infrared range based on metamaterial perfect absorbers. We demonstrate that metamaterial emitters not only achieve high emissivity near the ideal blackbody limit, but are also capable of engineering the emissivity over large bandwidth in a desired wavelength dependent manner.

The unit cell of a single band infrared metamaterial absorber is shown in Fig. \ref{Fig1}(a)(c), and consists of two metallic elements; a cross shaped resonator and ground plane, with a dielectric layer spaced in between. Dimensions of the cross (in microns) are: a$_{1}$ = 3.2, l = 1.95, w$_{1}$ = 0.5. The thickness of both metallizations are 0.1 $\mu$m and thickness of the dielectric spacer is 0.2 $\mu$m. This three layer metamaterial couples to both the electric and magnetic components of incident electromagnetic waves and allows for minimization of the reflectance, at a certain frequency, by impedance matching to free space.~\cite{Liu10} The metallic ground plane is fabricated to be thick enough to prevent light transmission and therefore guarantees a narrow band absorber with high absorptivity. As the resonance frequency of the cross resonator is proportional to its length, we may easily scale the above described design to longer or shorter wavelengths. We may also construct metamaterials consisting of different sub-lattices~\cite{yuan,bingham}, with different resonator lengths corresponding to different absorption bands. Based on this idea we combine two different metamaterial perfect absorbers to form a bipartite checker board unit cell, as shown in Fig. \ref{Fig1}(b)(d). The dimensions are, (in microns): a$_{2}$ = 7, l$_{1}$ = 3.2, l$_{2}$ = 2.0, w$_{2}$=0.4. The thickness of metal layers in this dual-band absorber are 0.1 $\mu$m with a 0.3 $\mu$m dielectric layer in between.

\begin{figure}[!]
\begin{center}
\includegraphics[ width=3.25in,keepaspectratio=true]%
{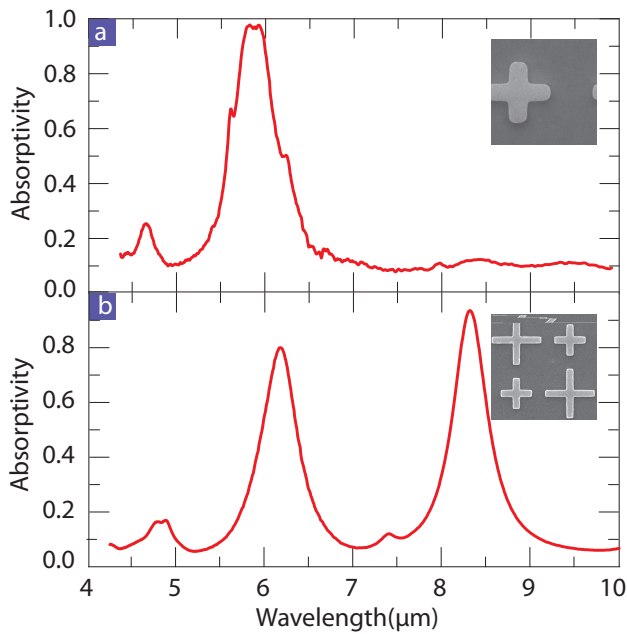}%
\caption{(Color online) a, Experimental absoprptivity of the single band metamaterial absorber. b, Experimental absoprptivity of the dual band metamaterial absorber. Inset displays SEM images of one unit cell for the fabricated single and dual band absorbers.}
\label{Fig2}%
\end{center}
\end{figure}

Fabrication of the single band metamaterial absorber begins with e-beam deposition of a 0.1 $\mu$m gold layer on top of a silicon substrate. A layer of benzocyclobutene (BCB) is then spin coated and cured on top of the gold layer and is a final thickness of 0.3 $\mu$m. The top cross resonator is fabricated with deep UV lithography and a photo mask. The use of deep UV lithography permits fabrication of large surface areas and here we realize 300 mm (12 inch) diameter samples where the fabricated metamaterial reticle has a total area of 4 mm $\times$ 4 mm. The dual band absorber is fabricated with e-beam lithography and, in this case, the dielectric spacer layer consists of Al$_2$O$_3$ deposited using atomic layer deposition (ALD) to a thickness of 0.2 $\mu$m. The total sample area of the dual band absorber is about 4.5 mm $\times$ 4.5 mm. Insets of Fig. \ref{Fig2} (a) and Fig. \ref{Fig2} (b) display SEM images of one unit cell for the fabricated single and dual band absorber, respectively.

We first characterized the absorptivity A($\omega$)=1-R($\omega$)-T($\omega$) for both the single and dual band metamaterials. Measurement of the reflection and transmission is performed with a Fourier-transform infrared microscope. A liquid-N2-cooled MCT is used for detection of radiation and is combined with a 15X cassegrain objective lens and KBr beam splitter. The reflection was measured at an incident angle of 20 degrees and transmission was measured at normal incidence. Reflection spectra are normalized with respect to a gold mirror and transmission with respect to an open aperture. The experimental absorptivity of the single band absorber is shown in Fig. \ref{Fig2} (a) and Fig. \ref{Fig2} (b) displays the absorptivity of the dual band absorber. We can see from Figs. \ref{Fig2} (a) and (b) that the single band absorber has an absorption peak at 5.8 $\mu$m with 97\% absorption and two absorption bands can be observed for the dual band absorber at 6.18 $\mu$m and 8.32 $\mu$m with 80\% and 93.5\% absorption, respectively.

\begin{figure}[!]
\begin{center}
\includegraphics[ width=3.5in,keepaspectratio=true]%
{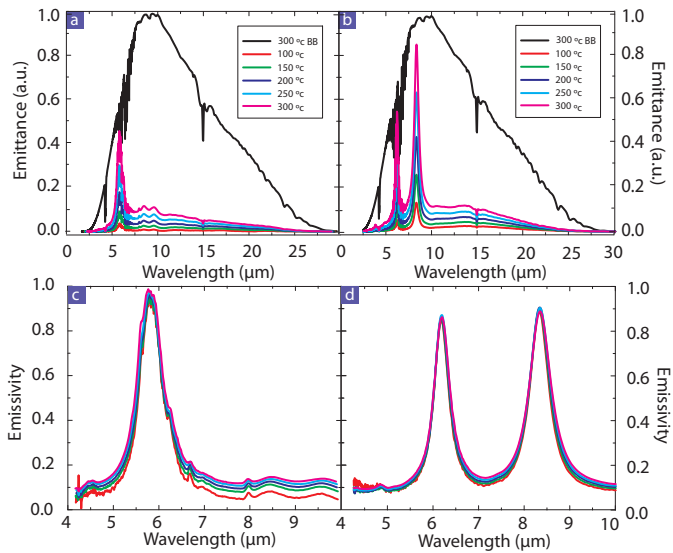}%
\caption{(Color online) Experimental emittance and normalized emissivity. a, Emittance of the single band metamaterial emitter at five different temperatures and Emittance of the blackbody reference at 300 $^\circ$C. b,Emittance of the dual band metamaterial emitter at five different temperatures and Emittance of the blackbody reference at 300 $^\circ$C. c, Normalized emissivity of single band metamaterial emitter. d, Normalized emissivity of dual band metamaterial emitter. }
\label{Fig3}%
\end{center}
\end{figure}

Experimental results presented in Fig.~\ref{Fig2} verify the ability of metamaterials for engineering absorption and we now turn toward experimental demonstration of tailored emissivity. An emission adaptor is attached to the external source port of the FT-IR system. A temperature controller permits emission studies from room temperature to 400 $^\circ$C in one degree steps. Radiation emitted from metamaterials is used as the source in the FT-IR spectrometer along with a KBr beam splitter and a DTGS detector. We characterize metamaterial emitters from 100 $^\circ$C to a maximum of 300 $^\circ$C in steps of 50 degrees.

In Fig. \ref{Fig3} (a) and (b) we display the measured emittance of the single and dual band metamaterial samples, respectively. For each a broad background is observed which increases for increasing temperature. Most notably, however, are sharp emittance peaks ranging from about 6 microns to just over 8 microns for both metamaterials. Our DTGS detector is not calibrated, thus emittances shown in Fig. \ref{Fig3} are in arbitrary units. We can, however, characterize the absolute value emissivity through characterization of the temperature dependent emittance of a blackbody reference - black carbon. Displayed as the black curve in Fig. \ref{Fig3} (a) and (b) is the blackbody reference emittance at 300 $^\circ$C. As can be observed, the emitted energy of the blackbody is significantly greater than that of the metamaterial emitters at the same temperature, except near 6 $\mu$m for the single band, and near 6 $\mu$m and 8 $\mu$m for the dual band sample, where metamaterial emissivities are nearly equal to that of the blackbody.

\begin{figure}[!]
\begin{center}
\includegraphics[ width=2.25in,keepaspectratio=true]%
{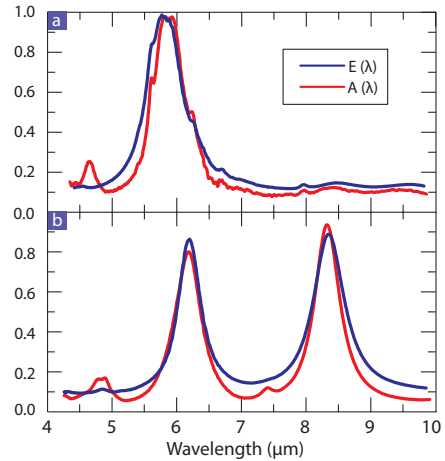}%
\caption{(Color online) Comparison between the experimental absorptivity and emissivity. a, absorptivity and emissivity for the single band absorber (emitter). b, absorptivity and emissivity for the dual band absorber (emitter).}
\label{Fig4}%
\end{center}
\end{figure}

We may calculate the absolute value metamaterial emittance by division with the emittance of a perfect blackbody, (at the same temperature), thus providing the emissivity. The metamaterial temperature dependent emissivities are displayed in Fig.\ref{Fig3} (c) and (d) for the single and dual band emitters. The single band emitter yields relatively low values throughout the range displayed, but notably has an emission peak of 98\% at 5.8 $\mu$m. Similar behavior is observed for the dual band emitter, i.e. it has a relatively low, wavelength independent, emittance punctuated by two emission peaks of 85\% and 89\% at 6.18 $\mu$m and 8.32 $\mu$m, respectively. The emissivity for both emitters exhibits a bit of temperature dependence which we attribute to thermal expansion of the dielectric spacer layer.

As a direct measure of utility of metamaterials to tailor emissivities, in Fig. \ref{Fig4} we compare the experimental absorptivity (red curves) and emissivity (blue curves) at 300 $^\circ$C, for both the single (a) and dual band (b) emitters. According to Kirchoff's law of thermal radiation, at equilibrium these should be equal for all wavelengths, i.e. $A(\lambda)=E(\lambda)$. Although the absorptivity is measured through transmission and reflection and the emissivity is characterized directly, we find good agreement between the two curves, as can be observed in Fig. \ref{Fig4}. This verifies that the emissive properties of metamaterials can be tuned by the absorption, i.e. by modifying the transmissive and reflective properties.

As a final example of the extreme flexibility of metamaterials to provide engineered emissivities over a broad bandwidth, we computationally design a metamaterial for use with a thermophotovoltaic cell for energy harvesting applications. A TPV essentially converts radiated energy (usually infrared) to electrons, and consists of a thermal emitter and a photovoltaic cell. Efficiency of the TPV is governed by the Shockley-Queisser (SQ) limit, which is highly dependent upon the mismatch between incident photon energy and the band gap (E$_{BG}$) of the semiconductor. One proposed method to increase TPV efficiency is to use selective emitters, whose emissivity is high within the TPV cell's sensitive region and low outside it. As a measure of the efficiency of TPVs we focus on the External Quantum Efficiency (EQE) which is the percentage of photons hitting the semiconducting surface that will produce an electron-hole pair. As a specific example we use the EQE of a low energy semiconductor gallium antimonide (GaSb)~\cite{Sulima02}. Our goal is to design a metamaterial emittance which peaks at the bandgap energy of GaSb, is low for energies below E$_{BG}$, and follows the EQE in a \emph{wavelength dependent manner} for E$>$E$_{BG}$.

\begin{figure}[!]
\begin{center}
\includegraphics[ width=3in,keepaspectratio=true]%
{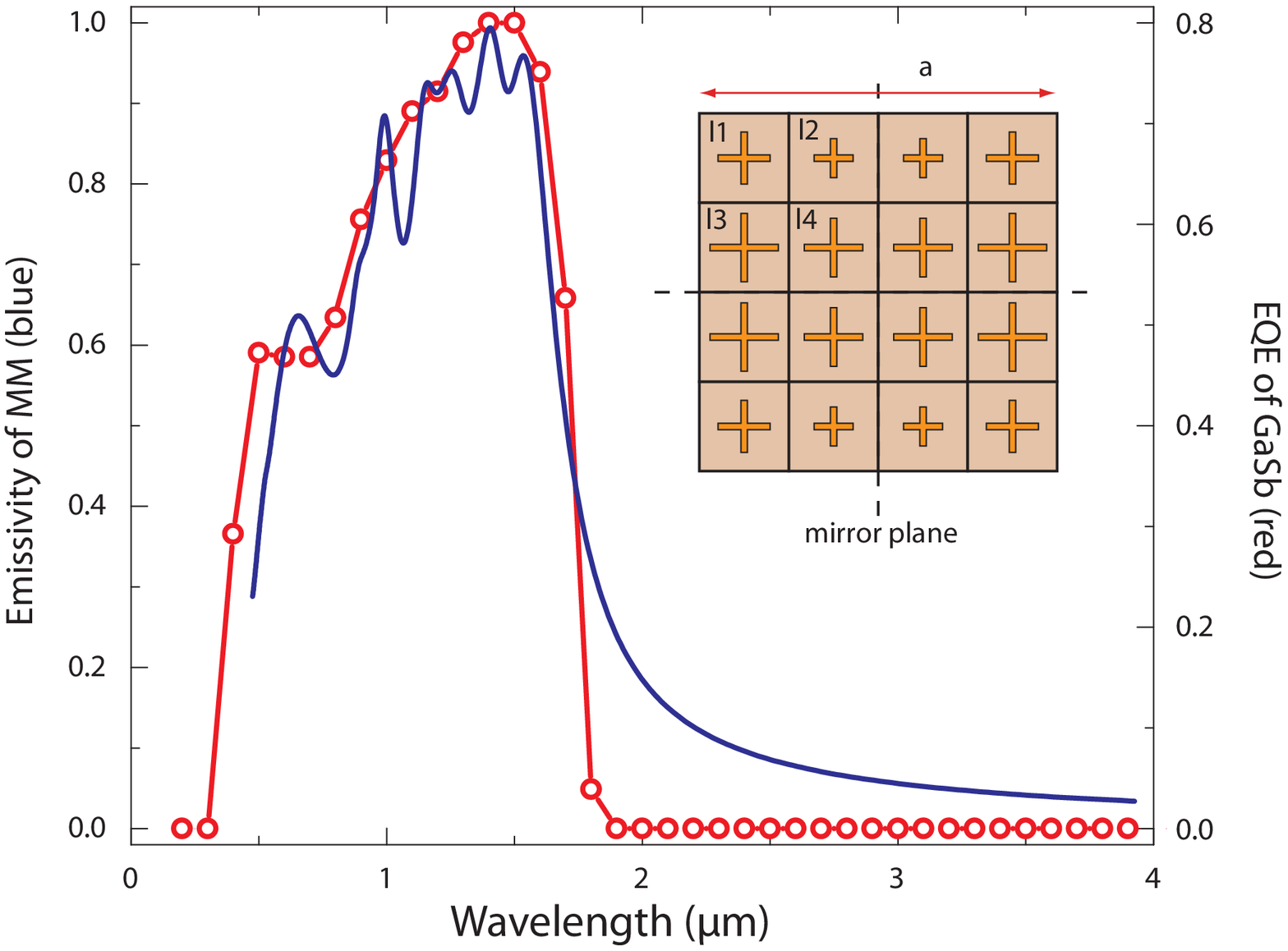}%
\caption{(Color online) Designed metamaterial emissivity (blue curve) to follow the EQE (red curve) of GaSb in a wavelength dependent fashion. Inset shows a schematic of the metamaterial design with sixteen sublattices. Dimensions of the design are a=1080nm, $1$=173nm, l$_2$=127nm, l$_3$=190nm, l$_4$=221.5nm. Each cross sublattice has a line width of 20nm and thickness of 50nm. The thickness of dielectric spacer is 78.8nm and ground plane is 200nm.}
\label{Fig5}%
\end{center}
\end{figure}

A metamaterial is computationally designed, following the same procedure previously outlined, and consists of a total of sixteen sublattices with dimensions of l$1$=173nm, l$_2$=127nm, l$_3$=190nm, l$_4$=221.5nm as shown in Fig. \ref{Fig5}. The size of the unit cell with 16 sublattices is a=1080nm and each cross sublattice has a line width of 20nm and thickness of 50nm. The dielectric spacer used in the simulation is Al$_2$O$_3$ with thickness of 78.8nm. Each sub-lattice consists of a cross resonator similar to those described for both the single and dual band emitters. As in the case of the dual band emitter, each sub-lattice independently achieves a designed emittance, both in terms of wavelength and amplitude. In this way we are able to engineer the emittance of the surface over a broad band, as shown in Fig. \ref{Fig5}. Here we have designed the emittance to follow the EQE of GaSb in a wavelength dependent fashion. As can be observed metamaterials permit extreme engineering of the emissivity of surfaces.

We have experimentally realized single and dual band metamaterial emitters. A comparison of emissivity to absorptivity shows good agreement, as predicted by Kirchoff's law of thermal radiation. Metamaterial emitters yield sharp bands (high Q) at infrared frequencies, peaking near the theoretical blackbody maximum. This suggests they are ideal candidates for energy harvesting applications. Other applications may benefit from the use of dynamical control~\cite{Chen08,Chen06,Chen09}, which may be implemented to provide tunable selective emitters which can be controlled by means of external stimuli. The utilization of polymer spacing layers combined with deep UV lithography implies their possible use for applications which require flexible large area control of emittance. A further benefit of the metamaterial emitters described here is their ability to achieve broad bandwidth response by the utilization of more complex sub-lattice designs. Through manipulation of the structures, these bands can be independently controlled in both magnitude and wavelength, thus providing precise control of thermal radiation in nearly any desired manner. The scale independent nature of metamaterials further suggests the option of engineering emissivities in other relevant bands of the electromagnetic spectrum.

\end{document}